\documentclass{article}

\usepackage{makeidx}         
\usepackage{graphicx}        
\usepackage{multicol}        
\usepackage{cite}            
\usepackage[bottom]{footmisc}

\usepackage{amssymb}

\usepackage[letterpaper,margin=1.6in]{geometry}

\makeindex             

\begin{document}

\title{Moment Closure - A Brief Review}

\author{Christian Kuehn\thanks{Institute for Analysis and Scientific Computing, 
Vienna University of Technology, 1040 Vienna, Austria}}


\maketitle

\begin{abstract}
Moment closure methods appear in myriad scientific disciplines in the modelling of complex 
systems. The goal is to achieve a closed form of a large, usually even infinite, set of 
coupled differential (or difference) equations. Each equation describes the 
evolution of one ``moment'', a suitable coarse-grained quantity computable from the full 
state space. If the system is too large for analytical and/or numerical methods, then one 
aims to reduce it by finding a moment closure relation expressing ``higher-order moments'' 
in terms of ``lower-order moments''. In this brief review, we focus on highlighting how 
moment closure methods occur in different contexts. We also conjecture via a geometric explanation 
why it has been difficult to rigorously justify many moment closure approximations although
they work very well in practice.
\end{abstract}

\section{Introduction}
\label{sec:intro}

The idea of moment-based methods is most easily explained in the context of stochastic
dynamical systems. Abstractly, such a system generates a time-indexed sequence of 
random variables $x=x(t)\in\mathcal{X}$, say for $t\in[0,+\infty)$ on a given state space 
$\mathcal{X}$. Let us assume that the random variable $x$ has a well-defined probability 
density function (PDF) $p=p(x,t)$. Instead of trying to study the full PDF, it is a natural 
step to just focus on certain moments $m_j=m_j(t)$ such as the mean, the variance, and so on, 
where $j\in\mathcal{J}$ and $\mathcal{J}$ is an index set and $\mathbb{M}=\{m_j:j\in\mathcal{J}\}$
is a fixed finite-dimensional space of moments. In principle, we may consider any moment space 
$\mathbb{M}$ consisting of a choice of coarse-grained variables approximating the full system, 
not just statistical moments. A typical moment-closure based study consists of four main steps:
\index{moment closure}

\begin{enumerate}
 \item[(S0)] \textbf{Moment Space:} \emph{Select} the space $\mathbb{M}$ containing a hierarchy 
 of moments $m_j$.
 \item[(S1)] \textbf{Moment Equations:} The next step is to derive evolution equations for the moments $m_j$. 
 In the general case, such a system will be \emph{high-dimensional} and fully \textit{coupled}.
 \item[(S2)] \textbf{Moment Closure:} The large, often even infinite-dimensional, system of moment
 equations has to be \emph{closed} to make it tractable for analytical and numerical techniques. In 
 the general case, the closed system will be \emph{nonlinear} and it will only \emph{approximate} the
 full system of all moments.
 \item[(S3)] \textbf{Justification \& Verification:} One has to justify, why the expansion made in step 
 (S1) and the approximation made in step (S2) are useful in the context of the problem considered. In 
 particular, the \emph{choice} of the $m_j$ and the \emph{approximation properties} of the closure
 have to be answered.
\end{enumerate}

Each of the steps (S0)-(S3) has its own difficulties. We shall not focus on (S0) as selecting 
what good 'moments' or 'coarse-grained' variables are creates its own set of problems. Instead,
we consider some classical choices. (S1) is frequently a lengthy computation. 
Deriving relatively small moment systems tends to be a manageable task. For larger systems, computer 
algebra packages may help to carry out some of the calculations. Finding a good closure in (S2) 
is very difficult. Different approaches have shown to be successful. The ideas frequently include 
heuristics, empirical/numerical observations, physical first-principle considerations or a-priori 
assumptions. This partially explains, why mathematically rigorous justifications in (S3) are 
relatively rare and usually work for specific systems only. However, comparisons with numerical 
simulations of particle/agent-based models and comparisons with explicit special solutions
have consistently shown that moment closure methods are an efficient tool. Here we shall also 
not consider (S3) in detail and refer the reader to suitable case studies in the literature.\medskip

Although moment closure ideas appear virtually across all quantitative scientific disciplines, a 
unifying theory has not emerged yet. In this review, several lines of research will be
highlighted. Frequently the focus of moment closure research is to optimize closure methods with one 
particular application in mind. It is the hope that highlighting common principles will eventually 
lead to a better global understanding of the area.\medskip

In Section~\ref{sec:equations} we introduce moment equations more formally. We show how to derive 
moment equations via three fundamental approaches. In Section~\ref{sec:closure} the basic ideas
for moment closure methods are outlined. The differences and similarities between different closure
ideas are discussed. In Section~\ref{sec:applications} a survey of different applications is given. 
As already emphasized in the title of this review, we do not aim to be exhaustive here but 
rather try to indicate the common ideas across the enormous breadth of the area.\medskip

\textbf{Acknowledgements:} I would like to thank the Austrian Academy of Science (\"OAW) 
for support via an APART Fellowship and the EU/REA for support via a Marie-Curie Integration
Re-Integration Grant. Support by the Collaborative Research Center 910 of the 
German Science Foundation (DFG) to attend the ``International Conference on Control of 
Self-Organizing Nonlinear Systems'' in 2014 is also gratefully acknowledged. Furthermore, I
would like to thank Thomas Christen, Thilo Gross, Thomas House and an anonymous referee for 
very helpful feedback on various preprint versions of this work.

\section{Moment Equations}
\label{sec:equations}

The derivation of moment equations will be explained in the context of three classical examples. 
Although the examples look quite different at first sight, we shall indicate how the procedures 
are related.

\subsection{Stochastic Differential Equations}
\label{ssec:SDE}

Consider a probability space $(\Omega,\mathcal{F},\mathbb{P})$ and let $W=W(t)\in\mathbb{R}^L$ be 
a vector of independent Brownian motions for $t\in\mathbb{R}$. A system of stochastic differential 
equations (SDEs) driven by $W(t)$ for unknowns $x=x(t)\in\mathbb{R}^N=\mathcal{X}$ is given by
\index{stochastic differential equation}
\begin{equation}
\label{eq:SDE0}
\textnormal{d} x = f(x)~\textnormal{d} t + F(x)~\textnormal{d} W
\end{equation}
where $f:\mathbb{R}^N\rightarrow \mathbb{R}^N$, $F:\mathbb{R}^N\rightarrow \mathbb{R}^{N\times L}$ 
are assumed to be sufficiently smooth maps, and we interpret the SDEs in the It\^o sense 
\cite{ArnoldSDEold,Kallenberg}. Alternatively, one may write (\ref{eq:SDE0}) using white noise, 
i.e., via the generalized derivative of Brownian motion\index{Brownian motion}\index{white noise}, 
$\xi:=W'$~\cite{ArnoldSDEold} as
\begin{equation}
\label{eq:SDE1}
x'=f(x)+F(x)\xi,\qquad '=\frac{\textnormal{d}}{\textnormal{d} t}.
\end{equation}
For the equivalent Stratonovich formulation see~\cite{Gardiner}. Instead of 
studying~(\ref{eq:SDE0})-(\ref{eq:SDE1}) directly, one frequently focuses on certain 
moments of the distribution. For example, one may make the \emph{choice} to consider
\begin{equation}
\label{eq:moments}
m_{\bf j}(t):=\langle x(t)^{\bf j}\rangle =\langle x_1(t)^{j_1}\cdots x_N(t)^{j_N}\rangle,
\end{equation}  
where $\langle \cdot\rangle$ denotes the expected (or mean) 
value\index{expected value}\index{mean} and ${\bf j}\in \mathcal{J}$, 
${\bf j}=(j_1,\ldots,j_N)$, $j_n\in\mathbb{N}_0$, where $\mathcal{J}$ is a certain set 
of multi-indices so that $\mathbb{M}=\{m_j:{\bf j}\in\mathcal{J}\}$. Of course, it should 
be noted that $\mathcal{J}$ can be potentially a very large set, e.g., for 
the cardinality of all multi-indices up to order $J$ we have
\begin{equation}
\nonumber
\left|\left\{{\bf j}\in\mathbb{N}_0^N:|{\bf j}|=\sum_n j_n\leq J\right\}\right|=
\left(\begin{array}{c}J+N\\ J \\\end{array}\right)=\frac{(J+N)!}{J!N!}.
\end{equation}
However, the main steps to derive evolution equations for $m_{\bf j}$ are 
similar for every fixed choice of $J,N$. After defining 
$m_{\bf j}=m_{\bf j}(t)$ (or any other ``coarse-grained'' variables), we may just 
differentiate $m_{\bf j}$. Consider as an example the case $N=1=L$, and 
$\mathcal{J}=\{1,2,\ldots,J\}$, where we write the multi-index simply as 
${\bf j}=j\in \mathbb{N}_0$. Then averaging~(\ref{eq:SDE1}) yields
\begin{equation}
\nonumber
m_1'=\langle x'\rangle =\langle f(x)\rangle +\langle F(x) \xi\rangle,
\end{equation} 
which illustrates the problem that we may never hope to express the moment 
equations\index{moment equation} explicitly for any nonlinear SDE if $f$ and/or $F$ are 
not expressible as convergent power series, i.e., if they are not analytic. The term 
$\langle F(x) \xi\rangle$ is not necessarily equal to zero for general nonlinearities $F$ as 
$\int_0^t F(x(s))~\textnormal{d} W(s)$ is only a \emph{local} martingale under relatively 
mild assumptions~\cite{Kallenberg}. Suppose we simplify the situation drastically 
by assuming a quadratic polynomial $f$ and constant additive noise\index{additive noise}  
\begin{equation}
\label{eq:simpleSDE}
f(x)=a_2x^2+a_1 x+a_0,\qquad F(x)\equiv \sigma\in\mathbb{R}.
\end{equation}
Then we can actually use that $\langle \xi\rangle=0$ and get
\begin{equation}
\nonumber
m_1'=\langle x'\rangle =a_2\langle x^2\rangle +a_1\langle x\rangle+a_0=a_2m_2+a_1m_1+a_0.
\end{equation}
Hence, we also need an equation for the moment $m_2$. Using It\^o's 
formula\index{It\^o's formula} one finds the differential
\begin{equation}
\nonumber
\textnormal{d} (x^2) = [2x f(x)+\sigma^2]~\textnormal{d} t + 2x \sigma~\textnormal{d} W
\end{equation}
and taking the expectation it follows that
\begin{eqnarray}
\nonumber
m_2'&=&2\langle a_2x^3+a_1 x^2+a_0 x\rangle +\sigma^2 + \sigma\langle 2x\xi \rangle \\
&=&2(a_2m_3+a_1m_2+a_0m_1)+\sigma^2,
\end{eqnarray}
where $\langle 2x\xi \rangle=0$ due to the martingale property\index{martingale} of 
$\int_0^t 2x(s)~\textnormal{d} W_s$. The key point is that the ODE for $m_2$ depends upon 
$m_3$. The same problem repeats for higher moments and we get an infinite system of 
ODEs, even for the simplified case considered here. For a generic nonlinear SDE, 
the moment system is a fully-coupled infinite-dimensional system of ODEs. Equations at a 
given order $|{\bf j}|=J$ depend upon higher-order moments $|{\bf j}|>J$, 
where $|{\bf j}|:=\sum_nj_n$.\medskip 

Another option to derive moment equations is to consider the Fokker-Plank (or forward 
Kolmogorov)\index{Fokker-Planck equation}\index{Kolmogorov equation} equation associated 
to (\ref{eq:SDE0})-(\ref{eq:SDE1}); see~\cite{Gardiner}. It 
describes the probability density\index{probability density} $p=p(x,t|x_0,t_0)$ of $x$ 
at time $t$ starting at $x_0=x(t_0)$ and is given by
\begin{equation}
\label{eq:FPE}
\frac{\partial p}{\partial t}=-\sum_{k=1}^N \frac{\partial}{\partial x_k}[pf]
+\frac12 \sum_{i,k=1}^N \frac{\partial^2}{\partial x_i \partial x_k}[(F F^T)_{ik}p].
\end{equation}
Consider the case of additive noise $F(x)\equiv \sigma$, quadratic polynomial 
nonlinearity $f(x)$ and $N=1=L$ as in~(\ref{eq:simpleSDE}), then we have
\begin{equation}
\label{eq:FPE1}
\frac{\partial p}{\partial t}=-\frac{\partial}{\partial x}[(a_2x^2+a_1 x+a_0)p ]
+\frac{\sigma^2}{2}\frac{\partial^2 p}{\partial x^2}.
\end{equation}
The idea to derive equations for $m_j$ is to multiply~(\ref{eq:FPE1}) by $x^j$, integrate by parts
and use some a-priori known properties or assumptions about $p$. For example, we have
\begin{eqnarray}
\nonumber
m_1'&=&\langle x'\rangle =\int_\mathbb{R} x\frac{\partial p}{\partial t}~\textnormal{d} x\\
\nonumber
&=&\int_\mathbb{R} 
-x\frac{\partial}{\partial x}[(a_2x^2+a_1 x+a_0)p ]~\textnormal{d} x +\int_\mathbb{R} 
x\frac{\sigma^2}{2}\frac{\partial^2 p}{\partial x^2}~\textnormal{d} x.
\end{eqnarray} 
If $p$ and its derivative vanish at infinity, which is quite reasonable for many densities, then 
integration by parts gives
\begin{equation}
\nonumber
m_1'=\int_\mathbb{R} [(a_2x^2+a_1 x+a_0)p ]~\textnormal{d} x = a_2m_2+a_1m_1+a_0
\end{equation} 
as expected. A similar calculation yields the equations for other moments. Using the 
forward Kolmogorov equation generalizes in a relatively straightforward way to
other Markov process, e.g., to discrete-time and/or discrete-space stochastic processes; in fact,  
many discrete stochastic processes have natural ODE 
limits~\cite{Kurtz,Kurtz1,DarlingNorris,BatkaiKissSikolyaSimon}. 
In the context of Markov processes, yet another approach is to utilize the moment generating 
function or Laplace transform $s\mapsto \langle \exp[\textnormal{i} sx]\rangle $ 
(where $\textnormal{i}:=\sqrt{-1}$)\index{moment generating function}\index{Laplace transform} 
to determine equations for the moments.

\subsection{Kinetic Equations}
\label{ssec:kinetic}

A different context where moment methods are used frequently is kinetic
theory~\cite{Levermore,Cercignani1,KrapivskyRednerBenNaim}. 
Let $x\in\Omega \subset \mathbb{R}^N$ and consider the description of a gas via a single-particle density 
$\varrho=\varrho(x,t,v)$, which is nonnegative and can be interpreted as a probability density if 
it is normalized; in fact, the notational similarity between $p$ from Section~\ref{ssec:SDE}
and the one-particle density $\varrho$ is deliberate. The pair $(x,v)\in\Omega\times \mathbb{R}^{N}$ 
is interpreted as position and velocity. A kinetic equation\index{kinetic equation} is given by
\begin{equation}
\label{eq:ke}
\frac{\partial \varrho}{\partial t}+v\cdot \nabla_x \varrho=Q(\varrho),
\end{equation}
where $\nabla_x=(\frac{\partial}{\partial x_1},\ldots,\frac{\partial}{\partial x_N})^\top$, suitable 
boundary conditions are assumed, and $\varrho\mapsto Q(\varrho)$ is the collision operator acting 
only on the $v$-variable at each $(x,t)\in\mathbb{R}^N \times [0,+\infty)$ with domain $\mathcal{D}(Q)$.
For example, for short-range interaction and hard-sphere collisions~\cite{MischlerMouhot} one would 
take for a function $v\mapsto G(v)$ the operator\index{collision operator}
\begin{equation}
\nonumber
Q(G)(v)=\int_{\mathbb{S}^{N-1}}\int_{\mathbb{R}^N}\|v-w\| [G(w^*)G(v^*)-G(v)G(w)]~\textnormal{d} 
w~\textnormal{d} \psi
\end{equation} 
where $v^*=\frac12(v+w+\|v-w\|\psi)$, $w^*=\frac12(v+w+\|v-w\|\psi)$ for $\psi\in\mathbb{S}^{N-1}$
and $\mathbb{S}^{N-1}$ denotes the unit sphere in $\mathbb{R}^N$. We denote velocity averaging by
\begin{equation}
\nonumber
\langle G\rangle =\int_{\mathbb{R}^N} G(v)~\textnormal{d} v,
\end{equation}
where the overloaded notation $\langle \cdot\rangle$ is again deliberately chosen to highlight the 
similarities with Section~\ref{ssec:SDE}. It is standard to make several assumptions about the 
collision operator such as the conservation of mass, momentum, energy as well as local entropy 
dissipation\index{entropy dissipation}
\begin{equation}
\label{eq:Qassume}
\langle Q(G)\rangle=0,\quad \langle v Q(G)\rangle=0,\quad \langle \|v\|^2Q(G)\rangle=0,\quad
\langle \ln(G)Q(G)\rangle \leq 0. 
\end{equation}
Moreover, one usually assumes that the steady states of~(\ref{eq:ke}) are 
Maxwellian\index{Maxwellian density} (Gaussian-like) densities of the form 
\begin{equation}
\label{eq:Maxwell}
\rho_*(v)=\frac{q}{(2\pi \theta)^{N/2}}\exp\left(-\frac{\|v-v_*\|^2}{2\theta}\right),\quad 
(q,\theta,v_*)\in\mathbb{R}^+\times \mathbb{R}^+\times \mathbb{R}^N
\end{equation}
and that $Q$ commutes with certain group actions~\cite{Levermore} implying symmetries. Note that the 
physical constraints~(\ref{eq:Qassume}) have important consequences, e.g., entropy dissipation implies
the local dissipation law
\begin{equation}
\label{eq:kecon2}
\frac{\partial}{\partial t}\langle \varrho\ln\varrho -\varrho\rangle +\nabla_x\cdot \langle 
v(\varrho\ln\varrho -\varrho) \rangle = \langle \ln \varrho Q(\varrho)\rangle \leq 0.
\end{equation}
while mass conservation implies the local conservation law\index{conservation law}
\begin{equation}
\label{eq:kecon1}
\frac{\partial}{\partial t}\langle \varrho\rangle +\nabla_x\cdot \langle v\varrho\rangle = 0
\end{equation}
with similar local conservation laws for momentum and energy. The local conservation law indicates
that it could be natural, similar to the SDE case above, to multiply the kinetic equation~(\ref{eq:ke})
by polynomials and then average. Let $\{m_j=m_j(v)\}_{j=1}^J$ be a basis for a $J$-dimensional 
space of polynomials $\mathbb{M}$. Consider a column vector $M=M(v)\in \mathbb{R}^J$ containing 
all the basis elements so that every element $m\in\mathbb{M}$ can be written as 
$m=\alpha^\top M$ for some vector $\alpha\in\mathbb{R}^J$. Then it follows    
\begin{equation}
\label{eq:ke1}
\frac{\partial }{\partial t}\langle \varrho M\rangle +\nabla_x \cdot \langle v\varrho M\rangle
=\langle Q(\varrho)M\rangle
\end{equation}
by multiplying and averaging. This is exactly the same procedure as for the forward
Kolmogorov equation for the SDE case above. Observe that~(\ref{eq:ke1}) is a $J$-dimensional
set of moment equations when viewed component-wise. This set is usually not closed.
We already see by looking at the case $M\equiv v$ that the second term in~(\ref{eq:ke1})
will usually generate higher-order moments.

\subsection{Networks}
\label{ssec:networks}

Another common situation where moment equations appear are network dynamical systems\index{networks}. Typical
examples occur in epidemiology, chemical reaction networks and socio-economic models. Here we 
illustrate the moment equations~\cite{Keeling,Rand,Tayloretal,SimonTaylorKiss} for the classical 
susceptible-infected-susceptible 
(SIS)\index{SIS model} model~\cite{DiekmannHeesterbeek} on a fixed network; for remarks on adaptive networks see 
Section~\ref{sec:applications}. Given a graph of $K$ nodes, each node can be in two states, 
infected $I$ or susceptible $S$. Along an $SI$-link infections occur at rate $\tau$ and recovery
of infected nodes occurs at rate $\gamma$. The entire (microscopic) description of the system is
then given by all potential configurations $x\in\mathbb{R}^N=\mathcal{X}$ of non-isomorphic graph 
configurations of $S$ and $I$ nodes. Even for small graphs, $N$ can be extremely large since already 
just all possible node configurations without considering the topology of the graph are $2^K$. Therefore,
it is natural to consider a coarse-grained description. Let $m_I=\langle I\rangle=\langle I\rangle(t)$ 
and $m_S=\langle S\rangle=\langle S\rangle(t)$ denote the average number of infected and susceptibles 
at time $t$. From the assumptions about infection and recovery rates we formally derive
\begin{eqnarray}
\frac{\textnormal{d} m_S}{\textnormal{d} t} & = & \gamma m_I - \tau \langle SI \rangle,
\label{eq:SI1}\\
\frac{\textnormal{d} m_I}{\textnormal{d} t} & = & \tau \langle SI \rangle - \gamma m_I,
\label{eq:SI2}
\end{eqnarray}      
where $\langle SI\rangle=:m_{SI}$ denotes the average number of $SI$-links. In~(\ref{eq:SI1}) the first 
term describes that susceptibles are gained proportional to the number of infected times 
the recovery rate $\gamma$. The second term describes that infections are expected to occur
proportional to the number of $SI$-links at the infection rate $\tau$. Equation~(\ref{eq:SI2})
can be motivated similarly. However, the system is not closed and we need an equation for
$\langle SI\rangle$. In addition to~(\ref{eq:SI1})-(\ref{eq:SI2}), the result~\cite[Thm.1]{Tayloretal}
states that the remaining second-order motif equations are given by
\begin{eqnarray}
\frac{\textnormal{d} m_{SI}}{\textnormal{d} t} & = & \gamma (m_{II} - m_{SI})
+\tau(m_{SSI} - m_{ISI} - m_{SI}),\label{eq:SI3}\\
\frac{\textnormal{d} m_{II}}{\textnormal{d} t} & = & -2\gamma m_{II} +2\tau(m_{ISI}+m_{SI}),\label{eq:SI4}\\
\frac{\textnormal{d} m_{SS}}{\textnormal{d} t} & = & 2\gamma m_{SI} - 2\tau m_{SSI}\label{eq:SI5},
\end{eqnarray}    
where we refer also to~\cite{Rand,Keeling}; it should be noted that~(\ref{eq:SI3})-(\ref{eq:SI5}) does not 
seem to coincide with a direct derivation by counting links~\cite[(9.2)-(9.3)]{DoGross}. In any case, 
it is clear that third-order motifs must appear, e.g., if we just look at the motif $ISI$ then an
infection event generates two new $II$-links so the higher-order topological motif structure does 
have an influence on lower-order densities. If we pick the second-order space of moments 
\begin{equation}
\label{eq:soepi}
\mathbb{M}=\{m_{I},m_{S},m_{SI},m_{SS},m_{II}\}
\end{equation}
the equations~(\ref{eq:SI1})-(\ref{eq:SI2}) and~(\ref{eq:SI3})-(\ref{eq:SI5}) are not closed. We 
have the same problems as for the SDE and kinetic cases discussed previously. The 
derivation of the SIS moment equations can be based upon formal microscopic balance 
considerations. Another option is write the discrete finite-size SIS-model as a Markov chain with 
Kolmogorov equation\index{Kolmogorov equation}
\begin{equation}
\label{eq:KSIS}
\frac{\textnormal{d} x}{\textnormal{d} t} = P x,
\end{equation}
which can be viewed as an ODE of $2^K$ equations given by a matrix $P$. One defines the 
moments as averages, e.g., taking
\begin{equation}
\nonumber
\langle I \rangle(t) :=\sum_{k=0}^K k x^{(k)}(t), \qquad \langle S \rangle(t) 
:=\sum_{k=0}^K (K-k) x^{(k)}(t),
\end{equation}
where $x^{(k)}(t)$ are all states with $k$ infected nodes at time $t$. Similarly one can
define higher moments, multiply the Kolmogorov equation by suitable terms, sum the equation 
as an analogy to the integration presented in Section~\ref{ssec:kinetic}, and derive the 
moment equations~\cite{Tayloretal}. For any general network dynamical systems, moment 
equations can usually be derived. However, the choice which moment (or coarse-grained) 
variables to consider is far from trivial as discussed in Section~\ref{sec:applications}. 

\section{Moment Closure}
\label{sec:closure}
\index{moment closure}
We have seen that moment equations, albeit being very intuitive, do suffer from the 
drawback that the number of moment equations tends to grow rapidly and the exact 
moment system tends to form an infinite-dimensional system given by
\begin{equation}
\label{eq:minf}
\begin{array}{lcl}
\frac{\textnormal{d} m_1}{\textnormal{d} t}&=&h_1(m_1,m_2,\ldots),\\
\frac{\textnormal{d} m_2}{\textnormal{d} t}&=&h_2(m_2,m_3,\ldots),\\
\frac{\textnormal{d} m_3}{\textnormal{d} t}&=&\cdots,\\
\end{array}
\end{equation}
where we are going to assume from now on the even more general case $h_j=h_j(m_1,m_2,m_3,\ldots)$
for all $j$.  In some cases, working with an infinite-dimensional system of moments
may already be preferable to the original problem. We do not discuss this direction
further and instead try to close~(\ref{eq:minf}) to obtain a finite-dimensional system. 
The idea is to find a mapping $H$, usually expressing the higher-order moments in terms of 
certain lower-order moments of the form
\begin{equation}
\label{eq:genclosuremap}
H(m_1,\ldots,m_\kappa)=(m_{\kappa+1},m_{\kappa+2},\ldots)
\end{equation}
for some $\kappa\in \mathcal{J}$, such that~(\ref{eq:minf}) yields a closed system 
\begin{equation}
\label{eq:minf1}
\begin{array}{ccc}
\frac{\textnormal{d} m_1}{\textnormal{d} t}&=&h_1(m_1,m_2,\ldots,m_\kappa,H(m_1,\ldots,m_\kappa)),\\
\frac{\textnormal{d} m_2}{\textnormal{d} t}&=&h_2(m_1,m_2,\ldots,m_\kappa,H(m_1,\ldots,m_\kappa)),\\
\vdots&=&\vdots\\
\frac{\textnormal{d} m_\kappa}{\textnormal{d} t}&=&h_\kappa(m_1,m_2,\ldots,m_\kappa,H(m_1,\ldots,m_\kappa)).\\
\end{array}
\end{equation}
The two main questions are 
\begin{enumerate}
 \item[(Q1)] How to find/select the mapping $H$? 
 \item[(Q2)] How well does~(\ref{eq:minf1}) approximate solutions of~(\ref{eq:minf}) and/or 
 of the original dynamical system from which the moment equations~(\ref{eq:minf}) have been derived? 
\end{enumerate}

Here we shall focus on describing the several answers proposed to (Q1). For a general nonlinear 
system, (Q2) is extremely difficult and Section~\ref{ssec:geometric_closure} provides a geometric
conjecture why this could be the case.

\subsection{Stochastic Closures}
\label{ssec:pdf_closure}
\index{stochastic closure}
In this section we focus on the SDE~(\ref{eq:SDE0}) from Section~\ref{ssec:SDE}. However, similar
principles apply to all incarnations of the moment equations we have discussed. One possibility is 
to \emph{truncate}\index{truncation} \cite{Socha} the system and neglect all moments higher than 
a certain order, which means taking
\begin{equation}
\label{eq:close1}
H(m_1,\ldots,m_\kappa)=(0,0,\ldots).
\end{equation}
Albeit being rather simple, the advantage of~(\ref{eq:close1}) is that it is trivial to
implement and does not work as badly as one may think at first sight for many examples. A 
variation of the theme is to use the method of steady-state of moments by setting 
\begin{equation}
\label{eq:QSS}
\begin{array}{ccc}
0&=&h_{\kappa+1}(m_1,m_2,\ldots,m_\kappa,m_{\kappa+1},\ldots),\\
0&=&h_{\kappa+2}(m_1,m_2,\ldots,m_\kappa,m_{\kappa+1},\ldots),\\
\vdots&=&\vdots\\
\end{array}
\end{equation}
and try to solve for all higher-order moments in terms of $(m_1,m_2,\ldots,m_\kappa)$ in the
algebraic equations~(\ref{eq:QSS}). As we shall point out in Section~\ref{ssec:geometric_closure}, this
is nothing but the quasi-steady-state\index{quasi-steady-state} assumption in disguise. 
Similar ideas as for zero and steady-sate moments can also be implemented using 
central moments and cumulants\index{cumulant}~\cite{Socha}.\medskip

Another common idea for moment closure principles is to make an \emph{a priori} assumption about 
the distribution of the solution. Consider the one-dimensional SDE example ($N=1=L$) and suppose
$x=x(t)$ is normally distributed. For a normal distribution\index{normal distribution} 
with mean zero and variance $\nu^2$, we know the moments
\begin{equation}
\langle x^j\rangle=\nu^j(j-1)!!,\quad \textnormal{if $j$ is even,}\qquad
\langle x^j\rangle= 0,\quad \textnormal{if $j$ is odd,}
\end{equation}
so one closure method, the so-called \emph{Gaussian (or normal) closure}\index{Gaussian closure}, is to set 
\begin{eqnarray}
\nonumber
 m_j&=&0\quad \textnormal{if $j\geq 3$ and $j$ is odd},\\
 \nonumber
 m_j&=&(m_2)^{j/2}~(j-1)!! \quad\textnormal{if $j\geq 4$ and $j$ is even.}
\end{eqnarray}
A similar approach can be implemented using central moments. If $x$
turns out to deviate substantially from a Gaussian distribution, then one has to question
whether a Gaussian closure is really a good choice. The Gaussian closure principle is one
choice of a wide variety of distributional closures. For example, one could assume the 
moments of a \emph{lognormal distribution}~\cite{EkanayakeAllen}\index{lognormal distribution} instead
\begin{equation}
\label{eq:logn}
x\sim \exp[\tilde\mu +\tilde\nu \tilde{x}],~\tilde{x}\sim\mathcal{N}(0,1),\quad \Rightarrow
\langle x^j\rangle=m_j=\exp\left[j\tilde\mu+\frac12 j^2\tilde\nu^2\right]
\end{equation}
where '$\sim$' means 'distributed according to' a given distribution and $\mathcal{N}(0,1)$ 
indicates the standard normal distribution. Solving for $(\tilde\mu,\tilde\nu)$ in~(\ref{eq:logn}) 
in terms of $(m_1,m_2)$ yields a moment closure $(m_3,m_4,\ldots)=H(m_1,m_2)$. The same
principle also works for discrete state space stochastic process, using a-prior 
distribution assumption. A typical example is the \emph{binomial closure}~\cite{KissSimon} 
and mixtures of different distributional closure\index{distributional closures} have also been 
considered~\cite{KrishnarajahCookMarionGibson,KrishnarajahMarionGibson}.

\subsection{Physical Principle Closures}
\label{ssec:entropy_closure}
\index{physical closure}
In the context of moment equations of the form~(\ref{eq:ke1}) derived from kinetic equations,
a typical moment closure technique is to consider a constrained closure based upon a postulated
physical principle. The constraints are usually derived from the original kinetic 
equation~(\ref{eq:ke}), e.g., if it satisfies certain symmetries, entropy dissipation and local 
conservation laws, then the closure for the moment equations should aim to capture these properties 
somehow. For example, the assumption 
\begin{equation}
\nonumber
\textnormal{span}\{1,v_1,\ldots,v_N,\|v\|^2\}\subset \mathbb{M}
\end{equation}
turns out to be necessary to recover conservation laws~\cite{Levermore}, while assuming that the 
space $\mathbb{M}$ is invariant under suitable transformations is going to preserve symmetries.
However, even by restricting the space of moments to preserve certain physical assumptions,
this usually does not constraint the moments enough to get a closure. Following~\cite{Levermore}
suppose that the single-particle density is given by 
\begin{equation}
\label{eq:expke}
\varrho=\mathfrak{M}(\alpha)=\exp[\alpha^\top M(v)],\qquad m=m(v)\in\mathbb{M}
\textnormal{ s.t. } m(v)=\alpha^\top M(v) 
\end{equation}
for some \emph{moment densities}\index{moment density} $\alpha=\alpha(x,t)\in\mathbb{R}^J$. Using~(\ref{eq:expke}) 
in~(\ref{eq:ke1}) leads to
\begin{equation}
\label{eq:ke2}
\frac{\partial }{\partial t}\langle \mathfrak{M}(\alpha)M\rangle +\nabla_x \cdot
\langle v\mathfrak{M}(\alpha) M\rangle=\langle Q(\mathfrak{M}(\alpha))M\rangle.
\end{equation}
Observe that we may view~(\ref{eq:ke2}) as a system of $J$ equations for the $J$ unknowns 
$\alpha$. Hence, one has formally achieved closure. The question is what really motivates
the exponential ansatz~(\ref{eq:expke}). Introduce new variables $\eta=\langle 
\mathfrak{M}(\alpha) M\rangle$ and define a function
\begin{equation}
\nonumber
H(\eta)=-\langle \mathfrak{M}(\alpha) \rangle+\alpha^\top \eta 
\end{equation}
and one may show that $\alpha=[\textnormal{D}_\eta H](\eta)$.
It turns out~\cite{Levermore} that $H(\eta)$ can be computed by solving the entropy 
minimization problem
\begin{equation}
\label{eq:entropy}
\min_\varrho \{\langle\varrho\ln\varrho -\varrho \rangle:\langle M\varrho\rangle =\eta \}=H(\eta),
\end{equation}
where the constraint $\langle M\varrho\rangle =\eta$ prescribes certain moments; we recall that 
$M=M(v)$ is the fixed vector containing the moment space basis elements and the relation 
$\alpha=[\textnormal{D}_\eta H](\eta)$ holds.
From a statistical physics perspective, it may be more natural to view~(\ref{eq:entropy})
as an entropy maximization\index{maximum entropy}\index{entropy} problem~\cite{Jaynes} by 
introducing another minus sign. Therefore, 
the choice of the exponential function in the ansatz~(\ref{eq:expke}) does not only guarantee
non-negativity but it was developed as it is the Legendre transform of the so-called 
entropy density $\varrho \mapsto \varrho \ln \varrho -\varrho$ so it naturally relates 
to a physical optimization problem~\cite{Levermore}.\medskip

To motivate further why using a closure motivated by entropy corresponds to certain
physical principles, let us consider the 'minimal'
moment space  
\begin{equation}
\nonumber
\mathbb{M}=\textnormal{span}\{1,v_1,\ldots,v_N,\|v\|^2\}
\end{equation}
The closure ansatz~(\ref{eq:expke}) can be facilitated using the vector $M(v)=(1,v_1,\ldots,v_N,\|v\|^2)$
but then~\cite{LevermoreMorokoff} the ansatz is related to the Maxwellian density~(\ref{eq:Maxwell}) 
since
\begin{equation}
\nonumber
\rho_*(v)=\exp[\alpha^\top M(v)],\quad \alpha=\left(\ln\left(\frac{q}{(2\pi\theta)^{3/2}}\right)
-\frac{\|v_*\|}{2\theta},\frac{v_*}{\theta},-\frac{1}{2\theta}\right)^\top
\end{equation}
but Maxwellian densities\index{Maxwellian density} are essentially Gaussian-like densities and 
we again have a \emph{Gaussian closure}. Using a Gaussian closure implies that the moment 
equations become the Euler equations\index{Euler equation} of gas dynamics, which can be viewed 
as a mean-field model near equilibrium for the mesoscopic single-particle kinetic equation~(\ref{eq:ke}), 
which is itself a limit of microscopic equations for each particle~\cite{Cercignani,Spohn2}.\medskip 

Taking a larger moment space $\mathbb{M}$ one may also get the Navier-Stokes equation as a 
limit~\cite{Levermore}, and this hydrodynamic limit can even be justified rigorously under 
certain assumptions~\cite{GolseSaintRaymond}. This clearly shows that moment closure methods 
can link physical theories at different scales.

\subsection{Microscopic Closures}
\label{ssec:micro_closure}
\index{microscopic closure}
Since there are limit connections between the microscopic level and macroscopic moment
equations, it seems plausible that starting from an individual-based network model, one
may motivate moment closure techniques. Here we shall illustrate this approach for the 
SIS-model from Section~\ref{ssec:networks}. Suppose we start at the level of first-order 
moments and let $\mathbb{M}=\{m_I,m_S\}$. To close~(\ref{eq:SI1})-(\ref{eq:SI2}) we want a map
\begin{equation}
m_{SI}=H(m_I,m_S).
\end{equation}
If we view the density of the $I$ nodes and $S$ nodes as very \emph{weakly correlated}\index{correlation} 
random variables then a first guess is to use the approximation
\begin{equation}
\label{eq:SISc1}
m_{SI}=\langle SI\rangle\approx \langle S \rangle \langle I\rangle=m_{S}m_{I}. 
\end{equation}
Plugging~(\ref{eq:SISc1}) into (\ref{eq:SI1})-(\ref{eq:SI2}) yields the \emph{mean-field}\index{mean-field} 
SIS model 
\begin{equation}
\label{eq:SIS_mf}
\begin{array}{lcl}
m_S' & = & \gamma m_I - \tau m_{S}m_{I},\\
m_I' & = & \tau m_{S}m_{I} - m_I.
\end{array}
\end{equation}     
The mean-field SIS model is one of the simplest examples where one clearly sees that although
the moment equations are \emph{linear} ODEs, the moment-closure ODEs are frequently 
\emph{nonlinear}. It is important to note that~(\ref{eq:SISc1}) is not expected to be valid for
all possible networks as it ignores the graph structure. A natural alternative is to consider
\begin{equation}
\label{eq:SISc1a}
m_{SI}=\langle SI\rangle\approx \mathfrak{m}_\textnormal{d}\langle S \rangle \langle I\rangle=
\mathfrak{m}_\textnormal{d} m_{S}m_{I}, 
\end{equation}
where $\mathfrak{m}_\textnormal{d}$ is the mean degree of the given graph/network. Hence it
is intuitive that~(\ref{eq:SISc1}) is valid for a complete graph in the limit 
$K\rightarrow \infty$~\cite{SimonTaylorKiss}.\medskip

If we want to find a closure similar to the approximation~(\ref{eq:SISc1}) for second-order
moments with $\mathcal{M}$ as in~(\ref{eq:soepi}), then the classical choice is the 
\emph{pair-approximation}~\cite{KeelingRandMorris,KeelingEames,GrossDLimaBlasius}\index{pair approximation}
\begin{equation}
\label{eq:SISc2}
m_{abc}\approx \frac{m_{ab}m_{bc}}{m_b},\qquad a,b,c\in\{S,I\}
\end{equation}
which just means that the density of triplet\index{triplet} motifs is given approximately by counting certain
link densities that form the triplet. In~(\ref{eq:SISc2}) we have again ignored pre-factors from the graph structure 
such as the mean excess degree~\cite{DoGross,Keeling}. As before, the assumption~(\ref{eq:SISc2}) is neglecting 
certain correlations and provides a mapping
\begin{equation}
\label{eq:SISc3}
(m_{SSI},m_{ISI})=H(m_{II},m_{SS},m_{SI})=\left(\frac{m_{SS}m_{SI}}{m_S},\frac{m_{SI}m_{SI}}{m_S}\right)
\end{equation}
and substituting~(\ref{eq:SISc3}) into~(\ref{eq:SI3})-(\ref{eq:SI5}) yields a system of five closed 
nonlinear ODEs. Many other paradigms for similar closures exist. The idea is to use the interpretation
of the moments and approximate certain higher-order moments based upon certain assumptions for
each moment/motif. In the cases discussed here, this means neglecting certain \emph{correlation terms}
from random variables. At least on a formal level, this is approach is related to the other closures
we have discussed. For example, forcing maximum entropy means minimizing correlations in the system
while assuming a certain distribution for the moments just means assuming a particular correlation
structure of mixed moments.

\subsection{Geometric Closure}
\label{ssec:geometric_closure}
\index{geometric closure}
All the moment closure methods described so far, have been extensively tested in many practical 
examples and frequently lead to very good results; see Section \ref{sec:applications}. However, 
regarding the question~(Q2) on approximation accuracy of moment closure, no completely general 
results are available. To make progress in this direction I conjecture that a high-potential 
direction is to consider moment closures in the context of geometric invariant manifold theory. 
There is very little mathematically rigorous work in this direction~\cite{StarkIannelliBaigent} 
although the relevance~\cite{DieckmannLaw1,PachecoTraulsenNowak} is almost obvious.\medskip 

Consider the abstract moment equations~(\ref{eq:minf}). Let us assume for illustration
purposes that we know that~(\ref{eq:minf}) can be written as a system 
\begin{equation}
\label{eq:minffs}
\begin{array}{ccc}
\frac{\textnormal{d} m_1}{\textnormal{d} t}&=&h_1(m_1,m_2,\ldots,m_\kappa,m_{\kappa+1},m_{\kappa+2},\ldots),\\
\frac{\textnormal{d} m_2}{\textnormal{d} t}&=&h_2(m_1,m_2,\ldots,m_\kappa,m_{\kappa+1},m_{\kappa+2},\ldots),\\
\vdots&=&\vdots\\
\frac{\textnormal{d} m_\kappa}{\textnormal{d} t}&=&h_\kappa(m_1,m_2,\ldots,m_\kappa,m_{\kappa+1},m_{\kappa+2},\ldots).\\
\frac{\textnormal{d} m_{\kappa+1}}{\textnormal{d} t}&=&\frac1\varepsilon h_{\kappa+1}(m_1,m_2,\ldots,
m_\kappa,m_{\kappa+1},m_{\kappa+2},\ldots).\\
\frac{\textnormal{d} m_{\kappa+2}}{\textnormal{d} t}&=&\frac1\varepsilon h_{\kappa+2}(m_1,m_2,\ldots,
m_\kappa,m_{\kappa+1},m_{\kappa+2},\ldots).\\
\vdots&=&\vdots\\
\end{array}
\end{equation}
where $0<\varepsilon\ll1$ is a small parameter and each of the component functions of the vector field 
$h$ is of order $\mathcal{O}(1)$ as $\varepsilon \rightarrow 0$. Then~(\ref{eq:minffs}) is a 
fast-slow system~\cite{KuehnBook,Jones}\index{fast-slow system} with fast variables 
$(m_{\kappa+1},m_{\kappa+2},\ldots)$ and slow 
variables $(m_1,\ldots,m_{\kappa})$. The classical \emph{quasi-steady-state}\index{quasi-steady-state} 
\emph{assumption}~\cite{SegelSlemrod} to reduce~(\ref{eq:minffs}) to a lower-dimensional system is 
to take
\begin{equation}
\nonumber
0=\frac{\textnormal{d} m_{\kappa+1}}{\textnormal{d} t},\qquad 0=\frac{\textnormal{d} 
m_{\kappa+2}}{\textnormal{d} t},\qquad\cdots.
\end{equation}
This generates a system of differential-algebraic equations\index{differential-algebraic equation} 
and if we can solve the algebraic equations 
\begin{equation}
\label{eq:alg}
0=h_{\kappa+1}(m_1,m_2,\ldots),\qquad 0=h_{\kappa+2}(m_1,m_2,\ldots),\qquad \cdots
\end{equation}
via a mapping $H$ as in~(\ref{eq:genclosuremap}) we end up with a closed system 
of the form~(\ref{eq:minf1}).\medskip 

The quasi-steady-state approach hides several difficulties that are best understood geometrically from the theory of 
normally hyperbolic invariant manifolds, which is well exemplified by the case of fast-slow 
systems. For fast-slow systems, the algebraic equations~(\ref{eq:alg}) provide a 
representation of the \emph{critical manifold}\index{critical manifold}
\begin{equation}
\nonumber
\mathcal{C}_0=\{(m_1,m_2,\ldots):h_j=0\textnormal{ for $j>\kappa$, $j\in\mathbb{N}$}\}.
\end{equation}
However, it is crucial to note that, despite its name, $\mathcal{C}_0$ is not necessarily a 
manifold but in general just an algebraic variety. Even if we assume that $\mathcal{C}_0$ is a 
manifold and we would be able to find a mapping $H$ of the form~(\ref{eq:genclosuremap}), this 
mapping is generically only possible \emph{locally}~\cite{Fenichel4,KuehnBook}. Even
if we assume in addition that the mapping is possible globally, then the dynamics on $\mathcal{C}_0$
given by~(\ref{eq:genclosuremap}) does not necessarily approximate the dynamics of the 
full moment system for $\varepsilon>0$. The relevant property to have a dynamical 
approximation is \emph{normal hyperbolicity}\index{normall hyperbolic}, i.e., the 'matrix'
\begin{equation}
\nonumber
\left.\left(\frac{\partial h_j}{\partial m_l}\right)\right|_{\mathcal{C}_0},\qquad 
j,l\in\{\kappa+1,\kappa+2,\ldots\}
\end{equation}
has no eigenvalues with zero real parts; in fact, this matrix is just the total derivative
of the fast variables restricted to points on $\mathcal{C}_0$ but for moment equations it is usually 
infinite-dimensional. Even if we assume in addition that $\mathcal{C}_0$ is normally hyperbolic, 
which is a very strong and \emph{non-generic} assumption for a fast-slow system~\cite{KuehnBook,Jones}, 
then the dynamics given via the map $H$ is only the \emph{lowest-order approximation}. The correct 
full dynamics is given on a \emph{slow manifold}\index{slow manifold}
\begin{equation}
\mathcal{C}_\varepsilon=\{(m_{\kappa+1},m_{\kappa+2},\ldots)=H(m_1,m_2,\ldots,m_\kappa)+
\mathcal{O}(\varepsilon)\}
\end{equation}
so $H$ is only correct up to order $\mathcal{O}(\varepsilon)$. This novel viewpoint on moment closure
shows why it is probably quite difficult~\cite{HasoferGrigoriu} to answer the approximation question 
(Q2) since for a general nonlinear system, the moment equations will only admit a closure via an explicit 
formula \emph{locally} in the phase space of moments. One has to be very lucky, and probably make very
effective use of special structures~\cite{PellisHouseKeeling,DelGenioHouse} in the dynamical system, to 
obtain any \emph{global} closure. Local closures are also an interesting direction to 
pursue~\cite{BoehmeGross}.

\section{Applications \& Further References}
\label{sec:applications}

Historically, applications of moment closure can at least be traced back to the classical Kirkwood 
closure~\cite{Kirkwood}\index{Kirkwood closure} as well as statistical physics applications, e.g., in the 
Ising model~\cite{Kikuchi}. 
The Gaussian (or normal) closure has a long history as well~\cite{Whittle}. In mechanical applications and
related nonlinear vibrations questions, stochastic mechanics models have been among the first where moment 
closure techniques for stochastic processes have become standard tools~\cite{Bolotin,Ibrahim} including the
idea to just discard higher-order moments~\cite{Richardson1}. By now, moment closure methods have permeated
practically all natural sciences as evidenced by the classical books~\cite{vanKampen,Adomian}. For SDEs, 
moment closure methods have not been used as intensively as one may guess but see~\cite{Bobryk}.\medskip 

For kinetic theory, closure methods also have a long history, particularly starting from the famous 
Grad 13-moment closure~\cite{Grad,StruchtrupTorrilhon}\index{Grad 13-moment closure}, and moment methods 
have become fundamental tools in gas dynamics~\cite{Struchtrup1}. One particularly important application 
for kinetic-theory moment methods is the modelling of plasmas~\cite{RobsonWhitePetrovic,HammettPerkins}. 
In general, it is quite difficult to study the resulting kinetic moment equations analytically~\cite{Desvillettes,Elmroth} 
but many numerical approaches exist~\cite{GrothMcDonald,LevermoreMorokoffNadiga,McDonaldGroth,TorrilhonStruchtrup}. 
Of course, the maximum entropy closure we have discussed is not restricted to kinetic theory~\cite{Singer1} 
and maximum entropy\index{maximum entropy} principles appear in many 
contexts~\cite{Abramov5,RanganCai,CernohorskyBludman,Csiszar,BorlandPenniniPlastinoPlastino}. \medskip 

One area where moment closure methods are employed a lot recently is mathematical biology. 
For example, the pair approximation~\cite{Keeling} and its variants~\cite{Bauch} are frequently 
used in various models including lattice 
models~\cite{SatoMatsudaSasaki,FilipeGibson,FilipeGibson1,Ellner,NakamarumatsudaIwasa,MatsudaOgitaSasakiSato}, 
homogeneous networks~\cite{PetermannDeLosRios,RozhnovaNunes} and many other network
models~\cite{BansalGrenfellMeyers,Rand2,Risau-GusmanZanette,VolzMeyers}. Several closures have also included 
higher-order moments~\cite{HouseDaviesDanonKeeling,Keeling1} and truncation ideas are still 
used~\cite{BolkerPacala,BolkerPacala1,HauskenMoxnes}. Applications to various different setups for epidemic
spreading are myriad~\cite{Hiebeler,HauskenMoxnes}\index{epidemics}. A typical benchmark problem for moment 
methods in biology is the stochastic logistic\index{stochastic logistic equation} 
equation~\cite{BartlettGowerLeslie,MatisKiffe,SinghHespanha,MatisKiffe1,Nasell,Nasell1,NewmanFerdyQuince}. 
Furthermore, spatial models in epidemiology and ecology have been a 
focus~\cite{OvaskainenCornell,LawDieckmann,Noeletal,MartchevaThiemeDhirasakdanon}. There are several 
survey and comparison papers with a focus on epidemics application and closure-methods 
available~\cite{BolkerPacalaLevin,MillerKiss,MurrellDieckmannLaw,Rand}. There is also a link from mathematical biology
and moment closure to transport and kinetic equations~\cite{Hillen,Hillen1}, e.g., in applications of
cell motion~\cite{Hillen2}. Also physical constraints, as we have discussed for abstract kinetic equations, 
play a key role in biology, e.g., trying to guarantee non-negativity~\cite{Hiebeler}.\medskip

Another direction is network\index{networks} dynamics~\cite{PorterGleeson}, where moment closure methods have been 
used very effectively are adaptive, or co-evolutionary, networks with dynamics of and on the 
network~\cite{GrossBlasius,GrossDLimaBlasius}. Moment equations are one reason why one may 
hope to describe self-organization of adaptive networks \cite{BornholdtRohlf} by low-dimensional 
dynamical systems models \cite{KuehnNetworks}. Applications include opinion 
formation~\cite{NardiniKozmaBarrat,KimuraHayakawa} with a focus on the classical voter 
model~\cite{SoodRedner,PuglieseCastellano,VazquezEguiluz}\index{voter model}; see~\cite{DemirelVazquezBoehmeGross} 
for a review of closure methods applied to the voter model. Other applications are found again 
in epidemiology~\cite{GrossKevrekidis,ShawSchwartz,ShawSchwartz1,TaylorTaylorKiss,Marceauetal,KuehnCT2,ZanetteRisau-Gusman} 
and in game theory~\cite{DemirelPrizakReddyGross,FuWuWang,DelGenioGross}. The maximum entropy-closure 
we introduced for kinetic equations has also been applied in the context of complex networks~\cite{Rogers} 
and spatial network models in biology~\cite{RaghibHillDieckmann}. An overview of the use of the 
pair approximation, several models, and the relation to master equations can be found in~\cite{Gleeson1}. 
It has also been shown that in many cases low-order or mean-field closures can still be quite 
effective~\cite{Gleesonetal}.\medskip 

On the level of moment equations in network science, one has to distinguish between purely moment
or motif-based choices of the space $\mathbb{M}$ and the recent proposal to use heterogeneous 
degree-based moments\index{heterogeneous}. For example, instead of just tracking the moment of a node 
density, one also characterizes the degree distribution~\cite{Gleeson} of the node via new moment 
variables~\cite{EamesKeeling}. Various applications of heterogeneous moment equations have been 
investigated~\cite{LindquistMaVandenDriesscheWilleboordse,SilkDemirelHomerGross}.\medskip

Another important applications are stochastic reaction 
networks~\cite{BarzelBiham,BarzelBiham1,GomezUribeVerghese}, where the mean-field reaction-rate 
equations are not accurate enough~\cite{LeeKimKim}. A detailed computation of moment equations from 
the master equation of reaction-rate models is given in~\cite{Engblom}. In a related area, 
turbulent combustion models are investigated using moment 
closure~\cite{Bilger,RoominaBilger,MortensenBilger,Klimenko,NavarroMartinezKronenbuergDiMare}.  
For turbulent combustion\index{turbulent combustion}, one frequently considers so-called conditional 
moment closures where one either conditions upon the flow being turbulent or restricts moments to 
certain parts of phase space; see \cite{KlimenkoBilger} for a very detailed review.\medskip

Further applications we have not focused on here can be found in genetics~\cite{BaakeHustedt}, 
client-server models in computer science~\cite{GuentherBradley,GuentherStefanekBradley}, mathematical 
finance~\cite{Singer}, systems biology~\cite{Gillespie}, estimating transport 
coefficients~\cite{ChristenKassubek}, neutron transport~\cite{BrunnerHolloway}, and radiative 
transport problems~\cite{FrankDubrocaKlar,Struchtrup}. We have also not focused on certain methods
to derive moment equations including moment-generating functions~\cite{Volz,HouseKeeling,Miller2}, 
Lie-algebraic methods~\cite{House}, and factorial moment expansions~\cite{Blaszczyszyn}.\medskip

In summary, it is clear that many different areas are actively using moment closure methods and
that a cross-disciplinary approach could yield new insights on the validity regimes of various 
methods. Furthermore, it is important to emphasize again that only a relatively
small snapshot of the current literature has been given in this review and a detailed account
of all applications of moment closure methods would probably fill many books.

{\small \bibliographystyle{plain}
\bibliography{../my_refs}}

%



\printindex
\end{document}